\documentclass{emulateapj}
\usepackage{graphicx}

\begin{document}

\title{Resolving the Origin of the Diffuse Soft X-ray Background}
\author{Randall K. Smith, Adam R. Foster, Richard J. Edgar, Nancy S. Brickhouse}
\affil{Harvard-Smithsonian Center for Astrophysics, 60 Garden Street, Cambridge MA 02138}

\begin{abstract}
The ubiquitous diffuse soft (1/4 keV) X-ray background was one of the earliest discoveries of X-ray astronomy. At least some of the emission may arise from charge exchange between solar wind ions and neutral atoms in the heliosphere, but no detailed models have been fit to the available data. Here we report on a new model for charge exchange in the solar wind, which when combined with a diffuse hot plasma component filling the Local Cavity provides a good fit to the only available high-resolution soft X-ray and extreme ultraviolet (EUV) spectra using plausible parameters for the solar wind.  The implied hot plasma component is in pressure equilibrium with the local cloud that surrounds the solar system, creating for the first time a self-consistent picture of the local interstellar medium.
\end{abstract}

\maketitle

\section{Introduction}
Early rocket flights using collimated proportional counters could easily detect and map soft X-ray emission in a range of bandpasses \citep{McS90}. These observations demonstrated that the emission must be local (coming from $< 100$\,pc) and eventually led to a picture that the emission was created by hot ($T_e \sim 10^6$K) plasma filling the Local Cavity surrounding the solar system \citep{McS90}, although it should be noted that this conclusion was based in large part by excluding all other mechanisms that could be imagined at the time.  The first measurement of high-resolution spectra of the soft X-ray background with the Diffuse X-ray Spectrometer (DXS) mission \citep{DXS01} was expected to confirm the conjecture of a thermal origin and shed light on specific creation mechanisms for the hot gas. As expected the spectrum was line-rich, but it did not match any existing models, including not only collisional ionization equilibrium models but also non-equilibrium models used, e.g. for supernova remnants \citep{SC01}.  Subsequent high-resolution EUV spectra of the diffuse emission with the Cosmic Hot Interstellar Plasma Spectrometer (CHIPS) \citep{CHIPS05} scanned the sky for months but failed to detect any of the predicted strong lines, confirming the failure of the original picture but shedding no light on a solution. 

The realization that X-rays emitted by comets \citep{Lisse96} arise from solar wind charge exchange (SWCX) \citep{Cravens97} led to the suggestion that this process could create at least part of the soft X-ray background \citep{Cox98}.  \citet{Cravens00} developed an analytic model that estimated the total X-ray flux from known solar wind ions interacting with interstellar neutral atoms flowing through the heliosphere, predicting that $\sim 50$\% of the soft X-ray background could be explained by SWCX.  More detailed models of the solar wind ions and the neutral H and He in the heliosphere \citep{Kout09} have extended this analysis, finding that charge exchange could contribute nearly all the flux in some directions. However, these models have suffered from incomplete atomic data, so that their predictions could not be directly compared to the high-resolution spectra. In addition, they do not explain the correlation of the surface brightness of the soft X-rays with the Local Cavity boundaries \citep{McS90}. 

We present a new approximate model for the charge exchange that includes a complete set of ions and radiative transitions and can be used in combination with other thermal models to fit the spectral emission observed by DXS and CHIPS.  Despite the limitations of the model, our results show for the first time that these data can be self-consistently fit with a combination of heliospheric SWCX combined with thermal emission, and that the parameters of both components are in agreement with other observations.  

\section{Method}

Our approximate approach does not involve detailed atomic models of the charge exchange emission process as a function of collision velocity and energy level, a challenging calculation that must be done for each ion individually.  Instead, the problem is separated into individual components.  Astrophysical X-ray charge exchange emission typically occurs when neutral hydrogen (or helium) loses an electron to a highly-ionized ion, usually into a high principal quantum number state which then decays radiatively. The total charge exchange emission along any line of sight thus depends upon the composition, density, and relative velocities of (i) the solar wind ions and (ii) the neutral material (either H or He), together with (iii) the cross section into each possible level and (iv) and the subsequent radiative cascade.  Existing models \citep{Cravens00,Kout07,Kout09} have focused on addressing (i,ii) while noting that many more atomic calculations (iii, iv) are needed.  These models and the observational data confirm that while the charge exchange component does vary, the total flux only changes by at most a factor of two to three and that predicting these variations is difficult.  

We chose to address issue (iv), the radiative cascade itself, with extensive new calculations for all relevant ions to get the best possible prediction for the emitted spectrum.  For all ions abundant in the solar wind, we extended the existing AtomDB \citep{Foster12} database of radiative rates and wavelengths with new calculations using the Autostructure \citep{badnell1986} code.  These new calculations extend the peak principal quantum number from a typical value of n=5 up to in some cases n=13.  
A key assumption of this model is that ions only undergo charge exchange, ignoring recombination and excitation by electrons. This is justified by the large cross section for charge exchange -- typically several orders of magnitude higher than the electronic cross sections that create most line emission in a collisionally excited plasma. \citet{janev1985} outline theoretical models for state selective charge exchange (SSCX) models, depending on the relative velocity of the ions $v$ and the characteristic orbital velocity $v_o = \sqrt{2I/m_e}$, where I is the ionization potential of the initial level of the donor electron (e.g. 13.6\,eV for ground state H, implying $v_o=2200$\,km/s), and $m_e$ is the electron mass. In these models, we assume that all the donor H$^0$ and He$^0$ ions are in the ground state, and therefore that the low velocity regime ($v/v_o \ll 1$) applies.  We assumed the heliospheric neutral component was 10\% He, as the DXS line of sight was at ecliptic latitude $\sim 90^{\circ}$\ (see \S \ref{sec:obs}) and did not intersect the He focusing cone \citep{Cravens00}.


The flux from CX is given by:
\begin{equation}
F_{CX} = {{1}\over{4\pi D^2}} \int{n_D n_R v \sigma^{cx}_{D\rightarrow R}(E) dV}
\end{equation}

where $F_{CX}$\ is the number of charge exchange photons, $n_D$\ and $n_R$\ are the donor and recombining ion densities, $v$\ is their relative velocity, and $\sigma^{cx}_{D\rightarrow R}(E)$ is the energy-dependent cross section for charge exchange between the donor and recombining ions. In the case of solar wind charge exchange (SWCX), the donor ions will be neutral hydrogen or helium atoms in the interstellar medium, while the recombining ions will be the solar wind ions. Of the four parameters determining the charge exchange rate, both $n_R$\ and $n_D$, and their variation along the line of sight are unknowns, while the relative velocities are known only approximately. Although some calculations exist, in general far too few velocity-dependent level-separated charge exchange cross section are available to populate a usable model. Instead, we chose to use relative cross sections, considering only the relative rates for capture in to different $n$ and $l$ shells using the methods described below, and use this to predict the shape of the charge exchange spectrum.  We obtain absolute magnitudes from fits to spectra and discretion during analysis.

\subsection{$n$-shell Distribution}
In the low velocity regime, the donor-receiver system can be modeled as a molecule, and the transfer of the electron as a resonant process. Following the arguments of \citet{janev1985}, the peak $n$ shell for capture, $n'$, is given by \citep[equation 2.6 of][]{janev1985}:
\begin{equation}
n'=q \sqrt{ \frac{1}{2}\frac{I_H}{I_p}} \left(1+\frac{q-1}{\sqrt{2q}}\right)^{-1/2}
\label{eq:ndist}
\end{equation}
where $I_H$\ is the 13.6 eV, the ionization energy of hydrogen, $I_p$\ is the ionization energy of the donor ion, and $q$\ is the charge of the ion receiving the electron.  At the low velocities we are interested in here, the capture into $n'$\ and then possibly one other adjacent $n$\ shell dwarfs the rate into other $n$ \citep{Ryufuku79}. For the purposes of this modeling we have assumed that all capture goes into the $n'$\ shell, or if $n'$\ is not an integer, then the fraction of the total capture going into the two nearest $n$\ shells, $f(n)$, is given by $f(n) = 1-|n'-n|$.

\subsection{Energy Level Matching}

Once $n'$\ has been calculated for an ion, all the possible LSJ coupled quantum states for addition of a single electron of $n \le n'$\ to the valence shell of the ground state configuration are identified. Energies and transition probabilities between these levels are taken from the AtomDB database version 2.0.2 \citep{Foster12} where they exist; for some ions, higher levels are required, and the energy levels and all dipole and quadrupole transitions between them are calculated using the \textsc{Autostructure} \citep{badnell1986} code. These are then merged with the existing AtomDB data, using the existing AtomDB preferentially when there is overlap.

\subsection{$l$-shell Distribution}
The emission from the captured electron will depend strongly on the distribution of angular momentum states populated by the CX process. Unfortunately, in the low energy regime the $l$-shell distribution is both broader than the $n$-shell distribution and more difficult to calculate, leading to a greater uncertainty \citep{Ryufuku79}.  In the absence of a detailed cross sections, \citet{janev1985} describe several methods for estimating the $l$ distribution, two of which provide analytic solutions in certain circumstances. We have implemented four different angular momentum distributions in this work:
\begin{itemize}
  \item{\textbf{Even:} each state $l$\ has an even weight}
  \item{\textbf{Statistical}: each level is weighted according to its statistical weight.}
  \item{\textbf{Landau-Zener}: each level is weighted by:
       \begin{equation}
       W(l) = l(l+1)(2l+1)\frac{(n-1)!(n-2)!}{(n+l)!(n-l-1)!}
       \end{equation}}
  \item{\textbf{Separable}: each level is weighted by:
       \begin{equation}
       W(l) = \left(\frac{2l+1}{q}\right)\exp\left[-\frac{l(l+1)}{q}\right]
       \end{equation}}
\end{itemize}
where $l$\ is the orbital angular momentum. 

{For reasons of convenience in our initial calculations we replaced the orbital angular momentum $l$\ with the total orbital angular momentum $L$. This was used for the fits presented here because there is no practical difference between the two for the Li-, He- and H-like ions which dominate X-ray spectra, and $L$\ is stored explicitly in AtomDB. Upon revisiting our calculations, 
we have added a new version which uses distributions based on the $nl$\ of the captured electron.  This latter approach better handles the intermediate weight ions where $LS$ coupling is inappropriate and $L$ is no longer a reliable quantum number, but presents a different set of problems with heavier ions where there is significant configuration mixing. For example, defining which levels truly represent a captured $11f$ electron is not exact. Given the qualitative nature of these distributions, we include both approaches in the distributed ACX model described below. Regardless of these differences, within all the levels which can be created by each capture with varying $L$, $S$, and $J$\ quantum numbers, we distribute the population using statistical weighting $(2J+1)$.}

\begin{figure}
\centering
\includegraphics[totalheight=2.5in]{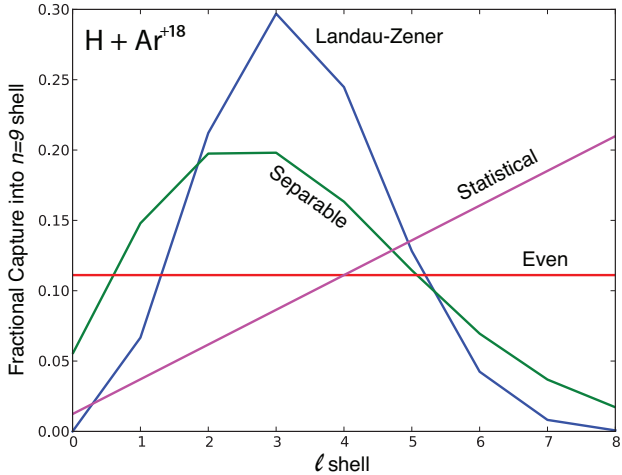}
\caption{\label{fig:compare_ldstn} The $l$ distribution of capture into the $n=9$ shell of the charge exchange reaction H + Ar$^{+18} \rightarrow$ H$^{+}$ + Ar$^{+17}$.}
\end{figure}

The four distributions here are not a comprehensive selection; they instead represent some likely and some extreme cases for comparison.  They are compared for capture into the $n=9$\ shell of Ar$^{+17}$\ in Figure~\ref{fig:compare_ldstn}. The Landau-Zener and Separable models are taken from Eqs 3.50a and 3.59 of \citet{janev1985}, and represent the very low velocity limit, with the peak $l$-shell being $l=1$\ or 2. As $v$\ increases, the population eventually becomes statistical. The even distribution was created to represent a situation somewhere in between these two extremes; it has no physical basis, but is included to help gauge the effect of the $l$-shell distribution on the resulting spectra.  {All of these cases, of course, ignore the well-known velocity-dependence of these distributions, at least at typical astrophysical velocities of less than a few thousand km s$^{-1}$, and so all of these models are merely approximate.}

In the case of the Statistical model, all the levels in a given $n$\ shell are populated completely statistically. For the Even, Landau-Zener and Separable models, the capture into $l$\ is determined by these models, but the split with the levels created by the $l$ electron is statistical. We note that these formulae are very approximate, and only applicable in highly ionized regimes. We have, however, applied them to all ions of all elements. While this is undoubtedly not ideal, state-selective charge exchange calculations for multi-electron ions are difficult, and rare. For many electron ions, such as Fe$^{+17}$, they are prohibitive. Therefore this model serves as a first approximation, which can be updated in future as better calculations or measurements become available. Most of the ions which contribute in the X-ray spectrum are one or two electron systems such as He or H-like metals, for which this formulation should be a fairly close approximation of reality.

\subsection{Spectrum Generation}

{
Having obtained the relative state-selective cross sections for capture, the calculation of the emitting spectrum follows easily. As stated earlier, the assumption is that every donor-ion pair undergoes charge exchange. We also assume the plasma is in the low electron density limit, so there are no collisional excitations during the cascade from the capture $n$\ shell to the ground. We have created two distinct models for different circumstances.  In the case of a dense source of neutral material such as a comet or a molecular cloud, all ions will undergo multiple charge exchange reactions as soon as the CX process begins. Thus a population of fully-ionized oxygen ions would undergo CX in this model to form hydrogen-like oxygen and emit one or more photon(s), and then these newly-created ions would again CX to form helium-like ions and emit another series of photons, continuing until neutral oxygen was reached.  This would be the appropriate model for a more distant charge exchanging plasma, such as the astrosphere of a star with its surrounding interstellar medium, where the entire interaction region is encompassed by the observations.
}

{
In the second model, however, each ion undergoes only a single charge exchange in the line of sight. Again assuming a population of fully-ionized oxygen, this model would assume CX formed hydrogen-like oxygen and emitted some photon(s), but no other ions would be included.  This we name the ``SWCX'' model, as it is intended to represent the solar wind charge exchange, and it is used for our analysis of the DXS and CHIPS data.  The solar wind is likely the only circumstance such a model should be used, because of the large cross sections involved.  Although rapid, even heliospheric CX has not fully neutralized the solar wind by 5 AU\,\citep{Cravens00}.  Unless carefully chosen to follow specific sightlines such as the He-focusing cone, a small field of view instrument will therefore be closer to this model than the `complete neutralization' approach described above.
}

{
As a result, the final model has several adjustable parameters. The $l$-distribution and the base/swcx model selection are two switches. The other key parameter is the ionization balance of the charge-exchanging plasma, which is used solely to determine the initial ionization state distribution of the recombining ions.  This ionization balance is assumed to be in collisional ionization equilibrium, which is parameterized by an electron temperature.  The donor ions can be a combination of both neutral hydrogen and helium, with the default value set to be their cosmic ratio.  This choice impacts the $n'$-shell after charge exchange, following Eq~\ref{eq:ndist}. Finally, the elemental abundances can be varied as well.  The final model is named ACX, the AtomDB Charge Exchange model, and has been coded into a form that can be used in XSPEC\citep{arnaud1996}\footnote{ACX is available at \url{http://www.atomdb.org/CX/}.}
}

\section{Observations}\label{sec:obs}

{
A number of telescopes have observed the soft X-ray background, but we focus on two in particular: the Diffuse X-ray Spectrometer (DXS)  and the Cosmic Hot Interstellar Plasma Spectrometer (CHIPS).  These missions were specifically designed to observe the spectrum of soft diffuse EUV and X-ray photons and determine their origin.  We briefly describe each mission's observing characteristics here; full details are available in the primary DXS \citep{DXS01} and CHIPS \citep{CHIPS05} papers.
}

{
The DXS experiment flew on the STS-54 mission (January 13-19, 1993), mounted in the Space Shuttle payload bay. The detector used a pair of rocking Bragg-crystal spectrometers to obtain spectra of the diffuse X-ray background in the $44-84$\AA (148-284 eV) range, with good $\sim 2.2$\AA\ spectral resolution but limited $\sim 15^{\circ}$\ angular resolution.  The observations were taken during the week of 1993 January 13, during orbit night.  The unobscured field of view is 15 degrees wide by $\sim 137$ degrees long, from Galactic longitude 160 to 297, and from Galactic latitude of about 15 degrees to 5, respectively. To avoid contamination from known supernova remnants (SNR), we excluded the MonoGem Ring and the Vela SNR.
}

{
The DXS viewing geometry relative to the heliosphere is complex, and so we discuss the expected signal from solar wind charge exchange from both geocoronal neutrals orbiting the Earth and interstellar neutrals in the heliosphere.  ROSAT observations of other solar system objects \citep{Dennerl10} show that the solar wind charge exchange emission tends to be roughly hemispherical, on the sunlit side. This is because the outer atmospheres of all planets and many comets are collisionally thick, that is, essentially every solar wind ion fully recombines through charge exchange reactions, emitting all the available X-rays on the sunlit hemisphere.  Since the DXS observations were taken from the night side of the Earth at Sun angles between 80 and 147 degrees, we expect little contamination of the signal from solar wind ions undergoing charge exchange with geocoronal neutrals as these are dominant only on the sunlit side.  The neutral component of the local interstellar medium streams through the heliosphere bound for a direction given by \citet{Mobius12} as (l,b) = (185.2, -12.0) degrees. The DXS look directions co-added for this analysis range from 34 to 105 degrees from this downwind direction. Therefore a large fraction of the heliospheric X-ray emission observed by DXS occurred along lines of sight at a large angle to the neutral flow, where the heliosphere should be collisionally thin to charge-exchange recombination.
}

{
The CHIPS satellite was launched in January 2003 and observed the EUV/soft X-ray background sky between 90-265\AA  with a peak resolution of $\sim 1.4$\AA for two years.  Like DXS, CHIPS had a relatively large $5^{\circ} \times 25^{\circ}$\ field of view, but it observed both during day and night and it covered a wide range of directions over the sky.  The designers expected to observe emission lines at or above 50 LU, with individual observations lasting 200 ksec to reach 3$\sigma$\ detections.  However, in practice only one strong line was observed (Fe IX at 171.1\AA) was observed, and all good observational data were combined into a single 13.2 Msec observation with overall upper limits of $\sim 6$\,LU. The CHIPS spectral data were not directly available to the authors in an easily usable form, but was not needed as the upper limits provided in \citet{CHIPS05} are adequate for this project.
}

\section{Results}

\begin{figure*}
\begin{center}
\includegraphics[totalheight=4in]{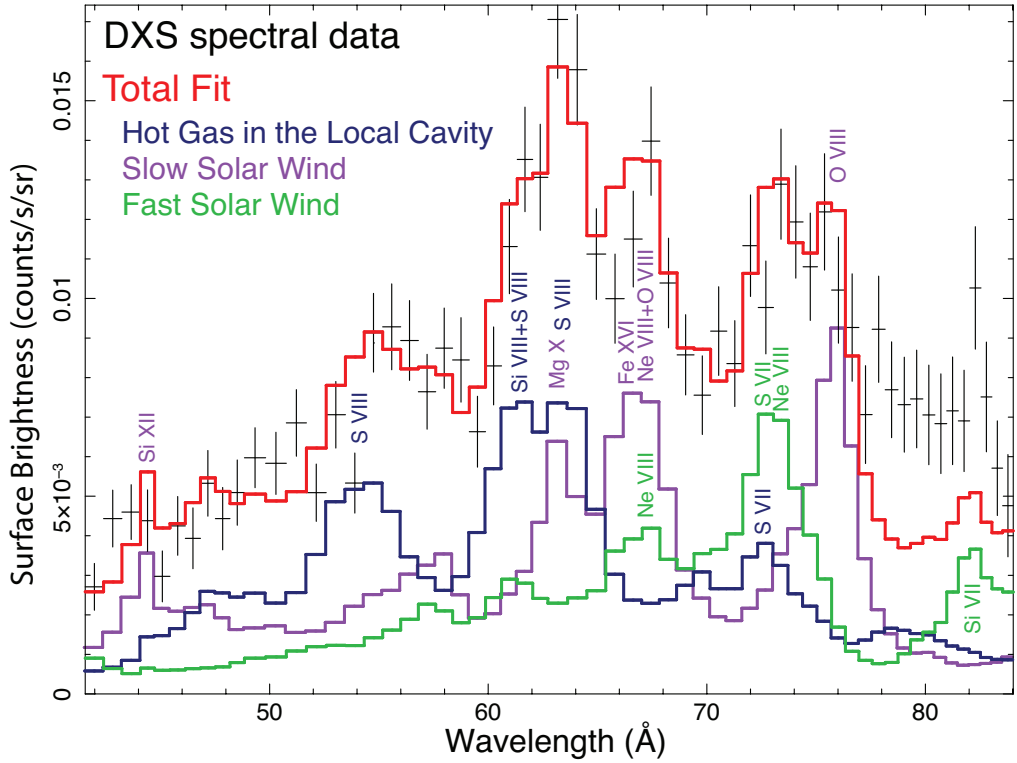}
\end{center}
\caption{The DXS data fit (red curve) using a combination of fast and slow solar wind charge exchange components (green, purple curves) plus a thermal component from hot plasma in the Local Cavity (blue curve); strong emission lines from each component are marked.  This fit also satisfies the upper limits set by the CHIPS mission in the 90-265\AA\ bandpass, although this drives the best-fit spectrum lower in the 76-84\AA\ range shown here. \label{fig:DXS}}
\end{figure*}

Figure~\ref{fig:DXS} shows the DXS spectrum of the diffuse soft X-ray background in the 1/4 keV bandpass, the highest resolution spectrum available.  The emission line features in this spectrum are unlike those of any thermal plasma model \citep{DXS01}. \citet{Wargelin04} tentatively identified the feature at $\sim 67.4$\AA\ as due to CX lines from Ne VIII and O VIII, identifications we confirm here. Although more detailed models that combine solar wind measurements with a partial set of charge exchange rates exist \citep{Cravens00,Kout07,Kout09}, they are not easily usable here since not all of the required solar wind conditions were monitored during the DXS measurements. The ACX model, in contrast, uses just the observed spectral data along with known solar parameters to determine the best-fit parameters.  The average measured abundance and ion population distributions of the solar wind \citep{vS00} constrain the model but the total charge exchange emissivity, which would depend upon the unknown details of the solar wind ions and the neutral material, is allowed to vary freely.  In addition to the DXS spectrum, we also consider the upper limits to the diffuse extreme ultraviolet ($90-265$\AA) emission measured by CHIPS \citep{CHIPS05}. 

We fit both the DXS and CHIPS datasets using the ACX model in combination with a component from a hot plasma filling the Local Cavity. This latter component is represented by a thermal plasma in collisional equilibrium \citep{Foster12} with abundances taken from those measured in nearby diffuse clouds \citep{SS96}. The charge exchange is modeled with interactions from both the slow and fast solar wind, using elemental abundances typical of each component \citep{vS00} with one modification, reducing the relative oxygen abundance by a third ({\it i.e.}\ 66\% of typical), as discussed below. 

Figure~\ref{fig:DXS} shows the best-fit model, fit using only six free parameters: the electron temperature for the collisional plasma model and two solar wind ionization populations, and three normalizations.  Since the electron density drops rapidly above the solar atmosphere, the ionization state of the solar wind is approximately ``frozen in'' in collisional ionization equilibrium at its source in the solar corona, reflecting the electron temperature of the hot plasma there. The temperature of the hot Local Cavity component ($1.12\pm0.06$\,MK) is in agreement with values predicted from magneto-hydrodynamic models \citep{SC01}, but creates only $26\pm4$\% of the total 0.1-0.4 keV flux. We also tested removing the hot thermal component and relying entirely on charge exchange.  This results in a significantly worse fit, with an F-test probability of $5\times10^{-10}$.  Using our best fit to the hot thermal component, the implied thermal pressure in the Local Cavity is $p/k_B = 5800 (100{\rm pc}/d)^{1/2}$\,cm$^{-3}$K, where $d$\ is the distance to the edge of the Cavity.  Unlike models of the soft X-ray emission that rely entirely on hot gas, which have pressures $p/k_B > 10,000$\,cm$^{-3}$K, our value is in equilibrium with the pressure in the Local Cloud surrounding the solar system \citep{FRS11}.  

\begin{figure*}
\begin{center}
\includegraphics[totalheight=2.3in]{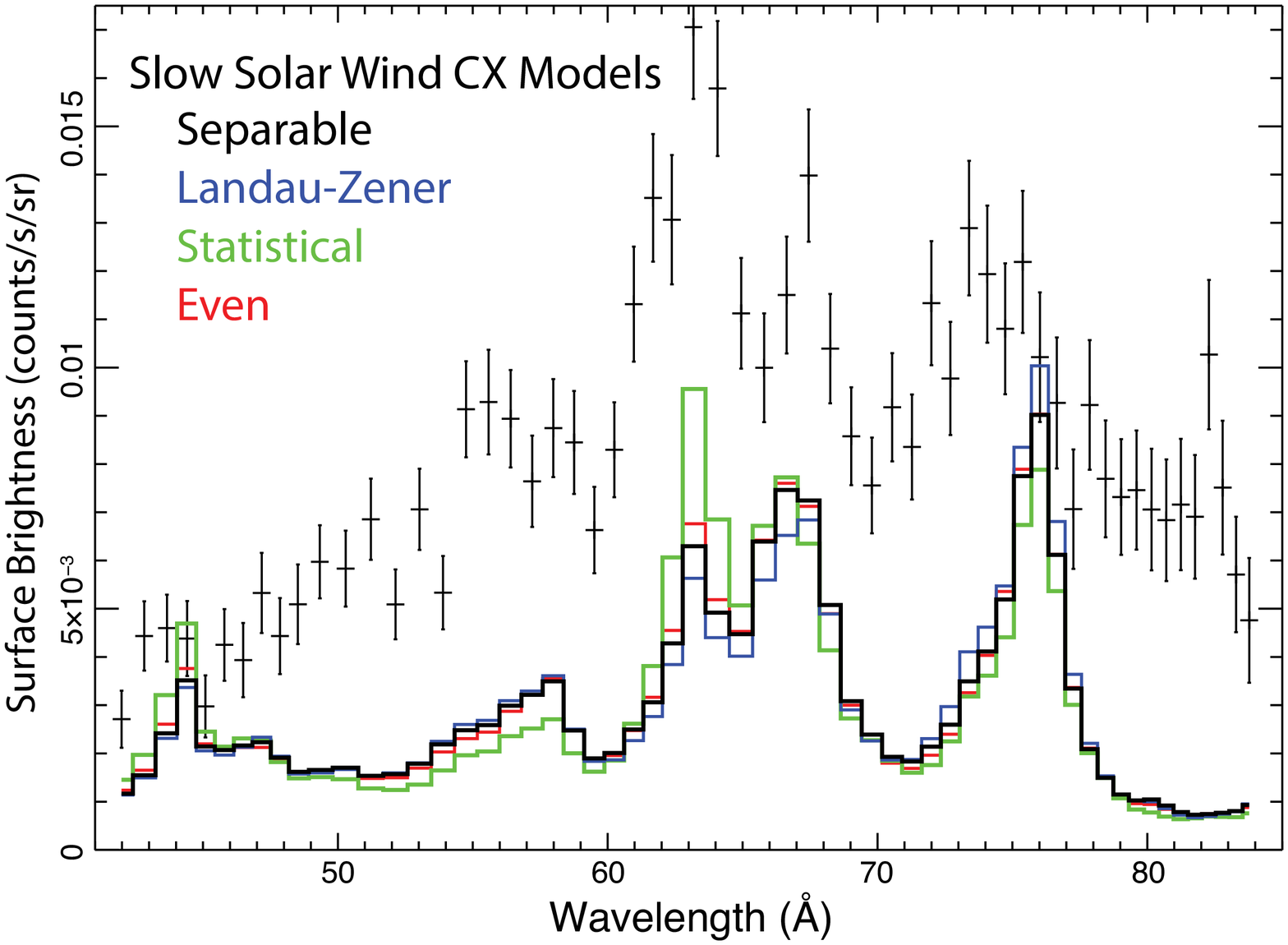}
\includegraphics[totalheight=2.3in]{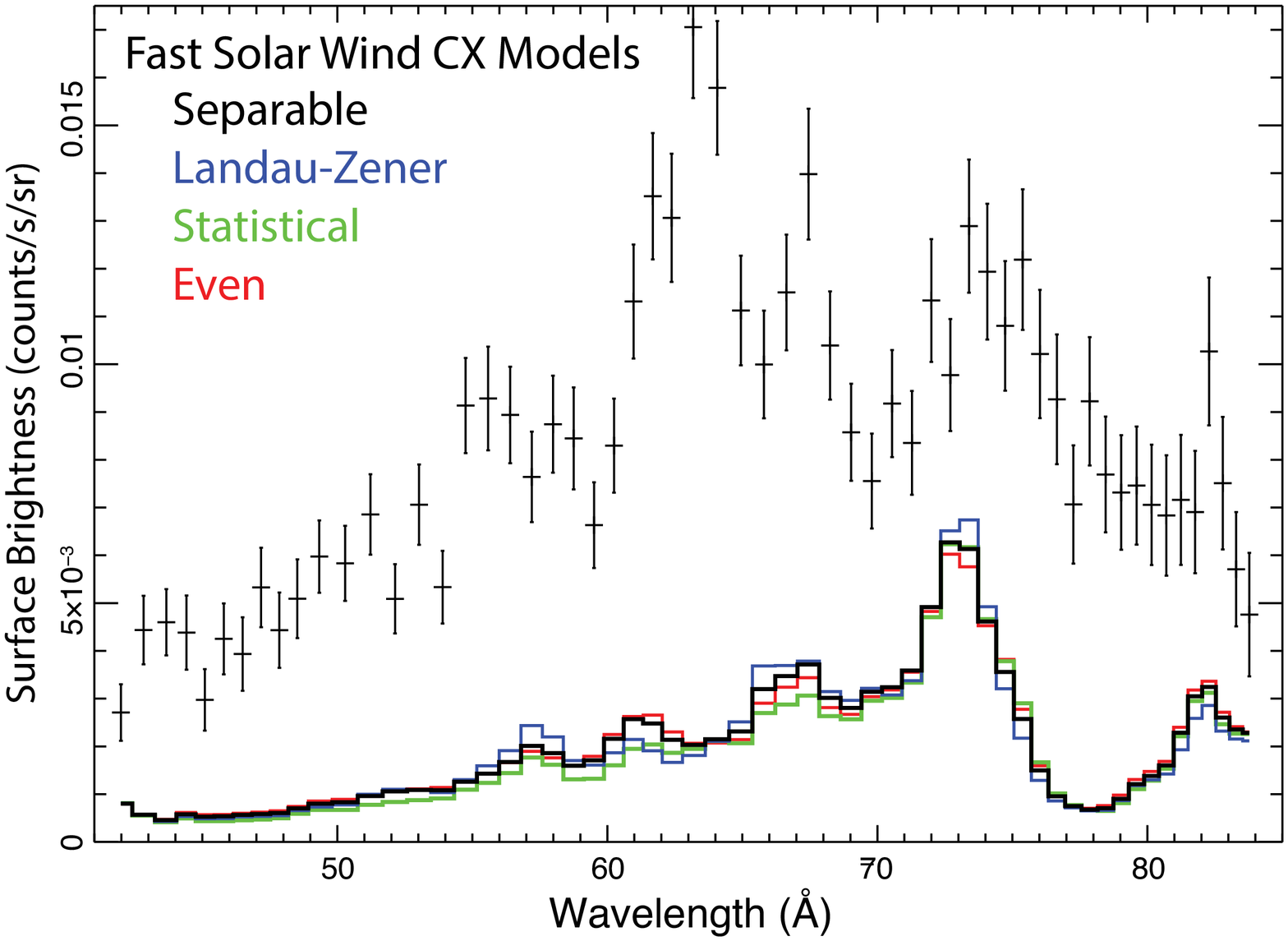}
\end{center}
\caption{The slow [Left] and fast [Right] solar wind charge exchange models comparing the four different $l$\ distribution models and all using our initial total angular momentum approach. The differences are greater in the slow solar wind, which is hotter and has more ion stages that emit in the X-ray band. Although not large to the eye, these variations cause slight but significant differences in the best-fit parameters. \label{fig:DXSothers}}
\end{figure*}

{
As noted above, we used the SWCX version of the model, along with the Separable approach to the $l$-shell distribution as the relative velocities are slow.  The Landau-Zener model would have been another reasonable choice, and in fact it gives similar (albeit slightly worse) fits.  The Even and Statistical distributions are not suitable for the velocities found in the solar wind and unsurprisingly give much poorer fits. For comparison, Figure~\ref{fig:DXSothers} shows models with the same parameters but changing the $l$-shell distribution used. 
}

The implied equilibrium frozen-in temperature for the fast solar wind is $0.85\pm0.1$\ MK, while that of the slow solar wind is $2.8^{+0.5}_{-0.3}$\,MK, in general agreement with existing observations, although the solar wind ions are not all drawn from a single temperature \citep{vS00}.  With only a single spectrum of modest resolution we cannot fit each element or ion independently, which would allow us to measure the ionic composition and abundances in the solar wind directly.  The DXS look direction is at $\sim 10^{\circ}$\ in Galactic latitude, a direction that must be dominated by local emission due to the large Galactic column density. The relative contributions of the charge exchange and hot plasma emission are in line with earlier detailed heliospheric models \citep{Kout09} that used a more limited selection of ions with, admittedly, more detailed $n,l$\ cross sections. Our model predicts only three observable features in the extreme ultraviolet band measured by CHIPS, whose combined 13.2 Msec spectrum has systematic background limits of $\sim 6$\,photons cm$^{-2}$s$^{-1}$sr$^{-1}$\,(Line Units, hereafter LU) \citep{CHIPS05}. These features include the Fe IX line at 171.1\AA (model 6.1 LU, observation $\sim 6$\,LU) and two O VI line complexes at 173\AA (7.2 LU) and 150\AA (6.6 LU). If present consistently throughout the CHIPS observations these lines might have been detected, but the DXS spectrum contains only $\sim 25$\,ksec of data and these lines may be variable. Other similar limits exist using different techniques; for example, the ALEXIS satellite used an imaging narrow-band filter that put $1\sigma$\ limits of $\sim 20$\,LU on any non-variable emission in the $\sim 170-185$\AA\ bandpass \citep{ALEXIS02}. We predict a constant flux in this bandpass of 17 LU from the hot plasma component plus a similar but variable amount from the charge exchange. This latter emission would have been variable and removed by the ALEXIS data analysis pipeline.

In addition to these EUV and soft X-ray observations, the Chandra, XMM-Newton, and Suzaku observatories have also measured diffuse soft X-rays at higher energies towards nearby dark clouds that shadow more distant emission, typically measuring emission from O VII ($\sim 22$\AA) of between 0.3-4.6 LU and O VIII ($\sim 19$\AA) of less than 2.1 LU \citep{Kout07}, albeit with large fluctuations likely caused by variations in the solar wind. The model predicts O VII emission of $\sim 4.2$\,LU and O VIII of $\sim 2.5$\,LU, within or slightly above the range of observed values \citep{Kout07}.  Although the DXS bandpass does not contain any strong oxygen lines, their presence both above and below this band strongly constrains the allowable models.  The surface brightness of these lines scales linearly with the assumed oxygen abundance, so reducing it in the solar wind models by a third remains within the one sigma limits of the abundance measurements \citep{vS00} while also remaining at or near the EUV and X-ray limits.

\section{Conclusions}

These results demonstrate that fitting the soft X-ray background requires both emission from charge exchange with the solar wind and hot gas within the Local Cavity, and that the CHIPS and DXS results can be simultaneously fit with a model consistent with the constraints from both solar system and Galactic measurements. The model does predict strong oxygen lines that push the oxygen abundance in the solar wind to the lower limits of the observed values, while preferring even lower values. Beyond explaining the origin and relative strengths of the components of the soft X-ray band emission, this result shows that long-term monitoring observations with good spectral resolution in this bandpass would allow indirect measurement of solar wind ions along any sightline, including at high ecliptic latitudes where direct measurements are extremely difficult. 

\acknowledgements{We thank John Raymond and Brad Wargelin for discussions and Jeffrey Morgenthaler for the original DXS data analysis and calibration, and most particularly the DXS PI Wilton Sanders, without whom this work would not have been possible.  This work was funded by NASA ADP NNX09AC71G and Chandra grant TM1-12004X. }

\end{document}